\documentclass[twoside]{dis07}
\usepackage[latin1]{inputenc}
\usepackage[dvips]{graphicx,epsfig,color}
\usepackage{wrapfig,rotating}
\usepackage{cite}
\usepackage{amssymb,amsmath,array}

\newcommand\Li{\,\mbox{${\rm Li}$}\,}

\newcommand\N{\nonumber}

\begin{document}
\title{{\normalsize\sl DESY 07/089     \hfill {\tt arXiv:yymm.nnnn}
\\ 
\vspace*{-2mm}
SFB/CPP-07-30 \hfill {   } }\\
Two-Loop Massive Operator Matrix Elements for Polarized and Unpolarized
Deep-Inelastic Scattering}

\author{I. Bierenbaum, J. Bl\"umlein and S. Klein 
%
\thanks{This paper was supported in part by SFB-TR-9: Computergest\"utze 
Theoretische Teilchenphysik, and the Studienstiftung des Deutschen 
Volkes.}
%
\vspace{.3cm}\\
%
Deutsches Elektronen-Synchrotron, DESY,
Platanenallee 6, D-15738 Zeuthen, Germany
}

\maketitle

\begin{abstract}
The $O(\alpha_s^2)$ massive operator matrix elements for unpolarized and 
polarized heavy flavor production at asymptotic values $Q^2 >> m^2$ are 
calculated in Mellin space without applying the integration-by-parts method. 
We confirm previous results given in Refs.~\cite{BU1,BU2}, however, obtain 
much more compact representations. 
\end{abstract}

\section{Introduction}

\noindent
The heavy-flavor corrections to deeply inelastic structure functions are 
very important for the range of small values of $x$ and do contribute there on the 
level of 20--40\%. They have to be known at the same level of accuracy as 
the light-flavor contributions for precision measurements of 
$\Lambda_{\rm 
QCD}$ \cite{SCALV} and the parton distributions. The next-to-leading 
order corrections were given semi-analytically in \cite{HQ} for the general 
kinematic range. Fast and  accurate implementations of these corrections 
in Mellin-space were given in \cite{AB}. In the region $Q^2 >> m^2$, the 
heavy flavor Wilson 
coefficients were  derived analytically to $O(\alpha_s^2)$ \cite{BU1,BU2}.
Here $Q^2$ denotes the virtuality of the gauge boson exchanged in deeply--inelastic 
scattering and $m$ is the mass of the heavy quark. In this note we summarize the results
of a first re-calculation of the operator matrix elements (OMEs) in \cite{BBK1,BK2}. The 
calculation is being 
performed in Mellin-space using harmonic sums \cite{HS1,HS2} without 
applying the 
integration-by-parts technique. In this way, we can significantly 
compactify both, the 
intermediary and final results. We agree with the results
in \cite{BU1,BU2}. The unpolarized and polarized $O(\alpha_s^2)$ massive OMEs can be used 
to calculate the asymptotic heavy-flavor Wilson coefficients 
for $F_2(x,Q^2)$ and $g_1(x,Q^2)$ to 
$O(\alpha_s^2)$~\cite{BU1,BU2,BBK1,BK2}, and for 
$F_L(x,Q^2)$  to $O(\alpha_s^3)$~\cite{FL}.  
\section{The Method}

\noindent
In the limit $Q^2 >> m^2$ the heavy quark contributions to the twist-2 Wilson 
coefficients are determined by {\sf universal} massive operator matrix elements
$\langle i|A_l|j \rangle$ between partonic states. The process dependence is due to
the corresponding massless Wilson coefficients \cite{WILS}.  This separation is obtained 
by applying
the renormalization group equation(s) to the (differential) scattering cross 
sections, cf.~\cite{BU1}. In this way all logarithmic and the constant contribution in
$m^2/Q^2$ can be determined. The operator matrix elements are calculated applying 
the operator insertions due to the light-cone expansion in the respective amplitudes.
One obtains the following representation
    \begin{eqnarray}
      {H_{(2,L),i}^{{\rm S,NS}}\left(\frac{Q^2}{\mu^2},
            \frac{m^2}{\mu^2}\right)} 
       = \underbrace{{A_{k,i}^{{\rm S,NS}} \left(\frac{m^2}{\mu^2}\right)}}_
	{
	\begin{array}{l}
	\mbox{massive OMEs}
	\end{array}
	} 
	\otimes \underbrace{{C_{(2,L),k}^{{\rm S,NS}} 
                \left(\frac{Q^2}{\mu^2}\right)}~,}_{
	\begin{array}{l}
          \mbox{light Wilson coefficients}
	\end{array}
        } \nonumber
    \end{eqnarray}
with $\otimes$ denoting the Mellin convolution. The OMEs contain 
ultraviolet and collinear divergences.
The collinear singularities are absorbed into the parton distribution 
functions while the ultraviolet divergences are removed through 
renormalization.
To 2--loop order, the renormalized 
OMEs read~:
\begin{eqnarray}
     {A_{Qg}^{(2)}} &=& \frac{1}{8}\left\{ {\widehat{P}_{qg}^{(0)}} 
                            \otimes 
                            \left[{P_{qq}^{(0)}} - {P_{gg}^{(0)}} 
                          + 2 \beta_0\right]\right\} {\ln^2\left(
                            \frac{m^2}{\mu^2}\right)}
                          - \frac{1}{2}{\widehat{P}_{qg}^{(1)}} 
                            {\ln \left(\frac{m^2}{\mu^2}\right)}\nonumber\\
                         & & +{\overline{a}_{Qg}^{(1)}} \otimes
                              \left[{P_{qq}^{(0)}}
                             -{P_{gg}^{(0)}} + 2 \beta_0\right]  
                             + {a_{Qg}^{(2)}}~, \nonumber
\end{eqnarray}
and similar for the quarkonic contributions. Here, $\mu^2$ denotes the 
factorization and
renormalization scale, $P_{ij}^{(k-1)}$ are the $k$th loop splitting 
functions and $\beta_0$ denotes the lowest expansion coefficient of the 
$\beta$--function.
${a}_{ij}^{(k)}$ and $\bar{a}_{ij}^{(k)}$ are the $O(\varepsilon^0)$ resp. 
$O(\varepsilon)$-terms in the expansion of the OME, which form the main 
objective
of the present calculation. 
\section{Results}

\noindent
We calculated the massive operator matrix elements both, for the 
gluon--heavy quark 
and light--heavy quark transitions in the flavor non-singlet and singlet 
cases, 
for unpolarized and polarized nucleon targets. 
 
The constant contribution to the unpolarized and polarized 
OMEs for the transition $g \rightarrow Q$ are~:

{\footnotesize   
  \begin{flalign}
   {a_{Qg}^{(2,{\rm unpol})}}(N)&=
        4 C_FT_R\Biggl\{
                     \frac{N^2+N+2}{N(N+1)(N+2)}
                       \Biggl[
                          -\frac{1}{3}{S_1^3(N-1)}
                          +\frac{4}{3}{S_3(N-1)} 
                          -{S_1(N-1)}{S_2(N-1)} \nonumber\\
                 &        -2{\zeta_2}{S_1(N-1)}
                        \Biggr]
                    +\frac{N^4+16N^3+15N^2-8N-4}
                          {N^2(N+1)^2(N+2)}{S_2(N-1)}
                    +\frac{3N^4+2N^3+3N^2-4N-4}
                          {2N^2(N+1)^2(N+2)}{\zeta_2} \nonumber\\
                 &  +\frac{2}{N(N+1)}{S_1^2(N-1)}
                    +\frac{N^4-N^3-16N^2+2N+4}
                          {N^2(N+1)^2(N+2)}{S_1(N-1)}
                    +\frac{P_1(N)}
                          {2N^4(N+1)^4(N+2)}
               \Biggr\}  \nonumber\\
        &+4C_A T_R\Biggl\{
                     \frac{N^2+N+2}{N(N+1)(N+2)} 
                       \Biggl[ 
                           4{{\rm M}~\Bigl[\frac{\Li_2(x)}{1+x}\Bigr](N+1)}
                          +\frac{1}{3}{S_1^3(N)}
                          +3{S_2(N)}{S_1(N)}\nonumber\\
                       &  +\frac{8}{3}{S_3(N)}
                          +{\beta''(N+1)} 
                          -4{\beta'(N+1)}{S_1(N)}
                          -4{\beta(N+1)}{\zeta_2} 
                          +{\zeta_3}
                       \Biggr]
                    -\frac{N^3+8N^2+11N+2}
                          {N(N+1)^2(N+2)^2}{S_1^2(N)}\N\\
                  & -2\frac{N^4-2N^3+5N^2+2N+2} 
                           {{(N-1)}N^2(N+1)^2(N+2)}{\zeta_2}
                    -\frac{7N^5+21N^4+13N^3+21N^2+18N+16} 
                          {{(N-1)}N^2(N+1)^2(N+2)^2} {S_2(N)}  \nonumber\\
                  & -\frac{N^6+8N^5+23N^4+54N^3+94N^2+72N+8}
                          {N(N+1)^3(N+2)^3}{S_1(N)}
                    -4\frac{N^2-N-4}
                           {(N+1)^2(N+2)^2}{\beta'(N+1)}\nonumber\\
                  & +\frac{P_2(N)}
                          {{(N-1)}N^4(N+1)^4(N+2)^4}\Biggr\}~.
\nonumber
     \end{flalign}
     \begin{flalign}
%
%
 {a_{Qg}^{(2,{\rm pol})}}(N)=&
           C_FT_R\Biggl\{
                        4\frac{N-1}{3N(N+1)}\Bigl(-4{S_3(N)}
                                                  +{S^3_1(N)}
                                                  +3{S_1(N)}{S_2(N)}
                                                  +6{S_1(N)}{\zeta_2} 
                                            \Bigr) \nonumber\\
&                       -4\frac{N^4+17N^3+43N^2+33N+2}
                                {N^2(N+1)^2(N+2)}{S_2(N)}
                         -4\frac{3N^2+3N-2}
                                {N^2(N+1)(N+2)}{S^2_1(N)} \nonumber\\
&                       -2\frac{(N-1)(3N^2+3N+2)}
                                {N^2(N+1)^2}{\zeta_2}
                         -4\frac{N^3-2N^2-22N-36}
                                {N^2(N+1)(N+2)}{S_1(N)}
                         -\frac{2P_3(N)}
                                {N^4(N+1)^4(N+2)}
                  \Biggr\} \nonumber\\
&         +C_AT_R\Biggl\{
                         4\frac{N-1}{3N(N+1)}\Bigl(
                             12{{\rm M}~\Bigl[\frac{\Li_2(x)}{1+x}\Bigr](N+1)}
                                    +3{\beta''(N+1)}
                                    -8{S_3(N)}
                                    -{S^3_1(N)} \N\\
&                                  -9{S_1(N)}{S_2(N)}
                                    -12{S_1(N)}{\beta'(N+1)}
                                    -12{\beta(N+1)}{\zeta_2}
                                    -3{\zeta_3}
                                            \Bigr)
                         -16\frac{N-1}
                                 {N(N+1)^2}{\beta'(N+1)} 
\nonumber\\
&                       +4\frac{N^2+4N+5}
                                {N(N+1)^2(N+2)}{S^2_1(N)} 
                         +4\frac{7N^3+24N^2+15N-16}
                                {N^2(N+1)^2(N+2)}{S_2(N)}
                         +8\frac{(N-1)(N+2)}
                                {N^2(N+1)^2}{\zeta_2} \nonumber\\
&                       +4\frac{N^4+4N^3-N^2-10N+2}
                                {N(N+1)^3(N+2)}{S_1(N)}
                         -\frac{4P_4(N)}
                                {N^4(N+1)^4(N+2)}
                  \Biggr\}~. \label{aqg2} \nonumber
     \end{flalign}
}

\vspace{0.5mm}\noindent
Here $P_i(N)$ denote polynomials given in \cite{BBK1,BK2}. 
The corresponding quarkonic expressions are given in \cite{BBK1,BK2}.
The integrals were performed
using Mellin-Barnes techniques \cite{BW,BBK2} and applying representations in terms of generalized 
hypergeometric 
functions. The results were further simplified using algebraic relations
between harmonic sums \cite{ALG}. Furthermore, structural relations for harmonic sums 
\cite{STRUCT}, which include half--integer relations and differentiation w.r.t. the Mellin 
variable $N$, lead to the observation that the OMEs above depend only on two basic harmonic 
sums~:
\begin{equation}
S_1(N), \hspace{2.5cm} S_{-2,1}(N)~.
\nonumber
\end{equation}
We expressed $S_{-2,1}(N)$ in terms of the Mellin transform $M[\Li_2(x)/(1+x)](N)$ in the
above.
Here $\beta(N) = (1/2) \cdot[\psi((N+1)/2) - \psi(N/2)]$.
Previous analyzes of various other space- and time-like 2--loop Wilson coefficients and 
anomalous dimensions including also the soft and virtual corrections to Bhabha-scattering
[15a,16], showed that six basic functions are needed in general to express 
these quantities~:
\begin{equation}
S_1(N), \hspace{4mm} S_{\pm 2,1}(N),\hspace{4mm} S_{-3,1}(N),\hspace{4mm} S_{\pm 2,1,1}(N)~. 
\nonumber
\end{equation}
None of the harmonic sums occurring contains an index $\{-1\}$ as observed 
in all other cases being analyzed. 

Comparing to the results  obtained in Refs.~\cite{BU1,BU2} in $x$--space, 
there 48 functions 
were needed to express the final result in the unpolarized case and 24 functions in the polarized case.

To obtain expressions for the heavy flavor contributions to the structure 
functions in $x$--space,
analytic continuations have to be performed to $N~\epsilon~{\bf C}$ for the basic functions given above,
see~\cite{STRUCT,AC1,AC2}. Finally a (numeric) contour integral has to be 
performed around the singularities
present. 
\section{Conclusions}

\noindent
We calculated the unpolarized and polarized massive operator matrix elements to $O(\alpha_s^2)$,
which are needed to express the heavy flavor Wilson coefficients contributing to the deep--inelastic
structure functions $F_2, g_1$ and $F_L$ to $O(\alpha_s^2)$ resp. $O(\alpha_s^3)$ in the region $Q^2 >> m^2$.
The calculation was performed in Mellin space without using the 
integration-by-parts technique, leading
to nested harmonic sums. We both applied representations through Mellin--Barnes integrals and generalized 
hypergeometric functions. In course of the calculations, a series of new 
infinite sums over products of 
harmonic sums weighted by related functions were evaluated, cf. \cite{BBK1,BK2}. These representations were 
essential to keep the complexity of the intermediary and final results as low as possible. Furthermore,
we applied a series of mathematic relations for the harmonic sums to compactify the results further.
We confirm the results obtained earlier in Refs.~\cite{BU1,BU2} by other technologies.

%
%
%
%
\begin{footnotesize} 

\end{footnotesize}
\end{document}